# Generative Blockchain: Transforming Blockchain from Transaction Recording to Transaction Generation through Proof-of-Merit

**Haozhao Zhang**
The Chinese University of Hong Kong, Shenzhen
**Zhe Zhang, Zhiqiang Zheng, Varghese Jacob**
University of Texas at Dallas, Naveen Jindal School of Management

## Abstract

This paper proposes a new paradigm: generative blockchain, which aims to transform conventional blockchain technology by combining transaction generation and recording, rather than focusing solely on transaction recording. Central to our design is a novel consensus mechanism, Proof-of-Merit (PoM), specifically crafted for environments where businesses must solve complex problems before transactions can be recorded. PoM integrates the generation and recording of transactions within a unified blockchain system, fundamentally differing from prevailing consensus mechanisms that primarily record existing transactions. We demonstrate PoM on a ride service on-demand platform, where the task of solving complex transaction-generating problems is delegated to a pool of independent problem solvers. These solvers generate transactions, and their solutions are selected based on merit. The winning solvers then register these transactions onto the blockchain and are rewarded accordingly. We introduce a Decentralized Control Parameter (DCP) to balance two key performance metrics: efficiency and equity. The applicability of our generative blockchain is illustrated through a ridesharing context, where matchers (solvers) are tasked with matching riders to drivers. We demonstrate PoM's performance and nuanced properties using agent-based simulation, exploring how to find the optimal DCP value to achieve a desirable balance of efficiency and equity in a generative blockchain.

**Keywords:** generative blockchain, Proof of Merit, transaction generation, agent-based simulation, decentralization

## 1. Introduction

In complex environments, businesses often face challenges in generating transactions before they can be recorded. For instance, Uber must match riders with drivers before a ride service transaction occurs, and the NBA must schedule games for its teams before any game-related transactions take place (Yang 2017). These challenges, often NP-hard tasks, directly influence the generation of business transactions. However, existing blockchain technologies primarily focus on validating and recording transactions (Ganne 2018), while transaction generation is kept outside the blockchain system.

This paper introduces a new paradigm: *generative blockchain*. Unlike traditional blockchain, which only records transactions, generative blockchain manages the generation of transactions within the blockchain itself. This approach reduces the computational load associated with solving transaction generation problems and the subsequent recording of these transactions. By integrating both processes, we streamline the workflow and optimize computing resource use, "killing two birds with one stone", efficiently manages both the generation and recording of transactions.

In addition to optimizing computing resources through the integration of generating and recording processes, another key advantage of our approach lies in decentralizing the generation of transactions. Traditionally, the initiation of business transactions often involves a centralized authority or system, which can be prone to bottlenecks, single points of failure, and potential security vulnerabilities. By incorporating a decentralized approach to the generation of transactions within the blockchain system, we introduce a layer of resilience and security to the overall transaction lifecycle. This decentralization distributes the responsibility of transaction generation across a network of participants, reducing dependency on a single entity and mitigating the risks associated with centralization. Each participant becomes a node contributing to the generation process, fostering a more robust system. This aligns with the fundamental principles of blockchain and contributes to a more trustworthy and efficient business transaction ecosystem.

Unlocking the seamless integration of transaction generating and recording hinges on the implementation of a novel consensus mechanism designed to harmonize both tasks within the blockchain system. We introduce a novel mechanism, Proof of Merit (PoM) that departs from conventional approaches that exclusively address post-generation processes and can only function after transactions have been created. PoM navigates the intricacies of both transaction generation and subsequent recording in a unified framework. Unlike traditional consensus mechanisms like Proof-of-Work (PoW), PoM replaces miners with solvers. A group of solvers undertakes the task of solving complex business problems before transaction generation. We leverage the fact that many of these complex problems, such as the generalized assignment problem, are NP-hard (Özbakir *et al*. 2010); therefore, the solvers will most likely need to use creative heuristics (Savelsbergh 1997; Zhang *et al*. 2020b; Cheng *et al*. 2021) to solve these problems The PoM mechanism selects winning solvers based on the merit of their solutions, acknowledging the varying qualities of solutions. The winning solutions are broadcast to users, and upon acceptance, transactions are realized (executed). Subsequently, the winning solver records these transactions onto the blockchain and receives a reward. All solvers can review the selected solutions and verify transactions through the blockchain. To illustrate the practical application of PoM, we choose the transportation service industry, specifically focusing on emerging technology in ride service on-demand platforms. We present a decentralized ridesharing platform where solvers, akin to matchers, play a crucial role in matching riders with drivers and seamlessly recording transactions onto the blockchain.

The overarching objective of PoM is to deliver high-quality solutions for complex transaction generation problems, while simultaneously maintaining a sufficient number of solvers, to ensure decentralization. We establish two key performance measures for PoM: (1) efficiency, as the degree to which the winning solution is close to the best solution known at that time; and (2) equity, as the degree of decentralization which is measured by the number of solvers participating in the solution-provisioning process. The diverse skill levels and various heuristics employed by solvers may introduce heterogeneity in the quality of solutions, potentially necessitating a tradeoff between efficiency and equity. For instance, consider a scenario where a few solvers consistently produce superior solutions compared to others. If selection is solely based on efficiency, these few solvers would dominate, limiting opportunities for other participants and potentially leading to their withdrawal (given their external options). While this approach maximizes efficiency, it comes at the expense of equity.

To strategically balance this tradeoff, we introduce a post-winning adjustment process in the PoM mechanism. This process incorporates a decentralization control parameter (DCP), mitigating the likelihood of a winner in one round repeating success in immediate subsequent rounds. This adjustment creates space for other solvers to succeed, thereby promoting a more balanced and decentralized participation in the solution-provisioning process.

Publicly available data for the validation and evaluation of our proposed Proof of Merit (PoM) is currently non-existent, given the absence of an operational blockchain application of this kind. Consequently, we turn to agent-based simulation (ABS) as a valuable tool for generating synthetic data to assess PoM's performance. Our comprehensive analysis involves measuring two crucial performance indicators, namely efficiency and equity, across various input conditions and decentralization control parameter (DCP) values. The simulation illuminates the counteracting effects of DCP; as DCP increases, equity improves at the cost of efficiency, and vice versa. Furthermore, we demonstrate that, contingent on specific input conditions such as the number of matchers, distribution of matcher capability, and acceptance rate from users being matched, the system designer can pinpoint a DCP level that attains desirable levels of equity and efficiency. This approach provides a nuanced understanding of PoM's performance in diverse scenarios, establishing a solid foundation for its potential application in real-world blockchain ecosystems.

The rest of the paper is structured as follows. We first provide a literature review in Section 2 and then formally construct the PoM consensus mechanism in Section 3. In Section 4, we discuss the agent-based simulation along with the insights on the performance of PoM under various conditions. Lastly, we conclude the paper with implications of the PoM design in Section 5.

## 2. Literature review

Blockchain, first utilized by Bitcoin, uses cryptography and a distributed ledger (DL) to enable a decentralized system in which the transactions are verified and securely recorded in a digital database that does not require control or maintenance by a centralized entity (*e.g.*, a firm). Instead, the DL is shared between and maintained by multiple individuals (nodes) in the network in a decentralized manner. Typically, new transactions are grouped and recorded in a “block” and chained to the previous blocks. Since its first application for Bitcoin, blockchains have been used in a myriad of applications.

A critical feature of blockchains is the consensus mechanism that is used to securely record transactions. Popular consensus mechanisms that include PoW and PoS have the goal of using them to reach a consensus - whereby a miner will be authorized to execute the task of adding a block of verified transactions to the blockchain. A series of studies (O'Dwyer and Malone, 2014, Eyal and Sirer, 2018) questioned whether PoW is sustainable due to the significant computing power being squandered. PoW is often criticized for (wastefully) consuming excessive computing power (Mishra *et al*. 2017). First used in Bitcoin, PoW requires miners to compete to solve a cryptographic hashing problem (the solution to which has no bearing on the content of the transaction to be recorded) and the winner compiles verified transactions into a new block and adds the block onto the Bitcoin blockchain. This design leads to hyper-competition among miners and unstainable energy wastage (Barrett 2022). In 2021, Bitcoin mining consumed more than 0.5% of all the electricity used globally and every single Bitcoin transaction consumed over $100 worth of electricity (Kim 2021, Tully 2021). To make matters worse, the computational resources used to solve the hashing problems provide no value to the users of Bitcoin beyond ensuring a secure and transparent transaction-recording process (Baldominos and Saez 2019). Under PoS, the miners with stakes in the system have the chance to be selected to register the new block onto the blockchain. However, PoS is often criticized for problems such as the “nothing at stake” issue, because there is no need for a block owner to spend any resource (Li *et al*. 2017), or “the rich get richer” issue where a small number of nodes could be repeatedly selected to create new blocks (Remus 2021). Wang *et al.* (2019) and Lashkari and Musilek (2021) provide a comprehensive survey on various consensus mechanisms used for blockchain. The Proof of Useful Work (PoUW) consensus mechanism, originally proposed by Ball *et al.* (2017), is closely related to PoM. It builds on PoW but replaces the energy inefficient hashing problems with solving complex but useful mathematical problems. The variants of PoUW differ in the external problems they

solve, e.g., finding complex optimization solutions, identifying gene sequence, and improving machine learning predictions (Shibata 2019, Turesson *et al*. 2021). Under PoUW, useful works are submitted by random (external) users and are unrelated to the transactions to be recorded on the blockchain.

In contrast, our proposed PoM mechanism has a unique objective: it is the first consensus mechanism that integrates the transaction generation and recording. In PoM, participants (problem solvers) solve the useful problems that directly determine how the transactions are to be generated and recorded. For example, in a rider-sharing context, solvers would propose solutions for rider-driver matching regarding which drivers should pick up which riders. The quality of the solutions (e.g., the total waiting time of riders) forms the basis to select the winning solver. Transactions are generated when the riders and the drivers accept the solution. The winner then earns the right to record the transactions onto the blockchain. Note that how to group transactions into a block and then record it to the blockchain is non-consequential to this study. For simplicity, we assume the winner groups all the realized transactions under her solution into a block and records the block onto the blockchain. Thus, PoM sets itself apart by targeting transaction generation – a considerably more complex problem than simply recording existing transactions, as done by traditional consensus mechanisms including PoW, PoS and PoUW. We use the term “transaction generation process” to refer to a two-step process: 1) solutions to the matching problems are generated and 2) based on the solutions, the transactions are realized, e.g., a rider and a driver accept the proposed match and fulfil the transaction.

Our paper also relates to the stream of incentive mechanism literature. Most studies in this literature stream (*e.g.*, Tong *et al*. 2018) have primarily focused on optimizing the profit of a centralized platform. Others, such as Yan *et al.* (2022), investigate the incentive mechanism for matching and pricing for drivers in a parking-sharing program, while Dong *et al.* (2024) propose a novel algorithm to solve the optimal scheduling problem on an on-demand service platform. While traditional consensus mechanisms such as PoW and PoUW approaches reward decentralized miners, they do not prevent these decentralized systems from collapsing into a centralized one due to the presence of a few dominant miners or stakeholders (Nasdaq 2020).

We also draw from the broader economics literature that explores the balance between efficiency and equity in various contexts, such as resource allocation (Okun 1975; Bardhan 1996; Tadenuma 2001; Badinelli 2010) and quality-based reward allocations (Winter and Watts 2010; Ho *et al*. 2016). Although the exact concepts of efficiency and equity vary in different contexts, the underlying ideas are the same: two substitutable objectives exist for evaluating social outcomes, such that attaining one objective may require sacrificing the other to some extent. Our study draws on these ideas but extends the literature to a decentralized service provisioning. We evaluate the effects of a key system design parameter (*i.e*., DCP) on the tradeoff between equity and efficiency. To the best of our knowledge, this is the first study integrating this tradeoff directly into the design of the consensus mechanism in a blockchain application.

We use agent-based simulation (ABS) to examine how key system design parameters influence PoM's performance. ABS has become a widely adopted method to conduct simulated experiments in Information Systems and other fields of study (Chan and Gao 2013; Zhang *et al*. 2020a). ABS serves a viable alternative for research questions that are difficult to be investigated through laboratory or field experiments, especially when controlling real-life factors can be infeasible (Zhang *et al*. 2020a). ABS enables researchers to precisely control the experiment parameters in a prescriptive manner, and thus provides an effective way to study the "what-if" question (Romme 2004). In the literature, it has been used to study the longitudinal dynamics in recommendation systems (Zhang *et al*. 2020a) and optimize regulation policy for government with unknown preferences (Yu *et al*. 2019).

ABS' prescriptive nature makes it especially suitable for studying mechanism design problems (Gregor 2006). In a typical ABS, the key design elements are agents, interactions, and an environment. Simulated agents (actors) can take actions to interact with each other and the environment based on some established rules (Amaral and Uzzi 2007). One needs to define the rules of the environment and various attributes of the agents (Nan 2011). In our implementation of ABS, we simulate a group of matchers offering solutions of various qualities, competing in each iteration governed by the PoM mechanism. The rule of the environment is set based on the winner selection process of the PoM consensus mechanism. Matchers interact with the environment by deciding whether to quit or stay in the system. And the simulation ends after the system reaches a point of steady state, where no matcher quits anymore.

## 3. PoM Consensus Mechanism

In this section, we describe our core design in building a generative blockchain - the PoM consensus mechanism. We note that PoM is generic and applicable to cases involving complex problem-solving, wherein solutions form the basis for transaction generation and recording. Specifically, PoM is suitable for contexts with the following characteristics: (1) transaction generation requires solving complex problems (*e.g.*, such as NP hard ones; Garey and Johnson 1990, Zhang *et al*. 2020b) for which an analytical (optimal) solution is hard to find; (2) a decentralized group of problem solvers contributes to the solution; (3) there exists a well-defined and commonly accepted metric to assess solution quality; and (4) end users have private information and may approve or disapprove the solutions. Because the problem-solving process involves numerous decentralized problem solvers, a blockchain is essential to store information (solutions, transactions, etc.) and crucially, to establish trust in such a decentralized setting. Next, we introduce the players with their characteristics and then define the task of problem solving.

### 3.1 Design of PoM

Imagine a group of users who post their problems on a platform. These problems must be solved cohesively before these users' transactions can be generated. The platform employs the PoM consensus mechanism to solicit and select solutions from solvers, who are autonomous agents using heuristic algorithms in deciding how users' problems should be solved.

We divide the timeline into multiple problem-solving rounds and each round is indexed by $t$, $t \in \{1,2, \dots, T\}$, where $T$ denotes the total number of rounds after which the system is naturally reset to the initial state. We use $R_t$ to represent the set of users in a matching round $t, t \in \{1,2, \dots, T\}$, and use $M_t$ to represent the set of solvers at $t$. To focus on the process of the PoM mechanism, we do not model explicitly how solvers solve users' problems; instead, we assume the solvers obtain heterogenous solutions as given.

We use $X_t$ to represent the set of solvers' solutions, where solver $k$'s solution is $X_t^k$, $k \in M_t$ at $t$ (*see* Table A1 in online supplement for the list of notations).

In each solving round, the system selects only one solution and then broadcasts the chosen one to all the users. The solver of the selected solution is rewarded for a fixed amount, which is exogenous and public information, whereas the remaining solvers receive zero reward, *i.e.*, winner-takes-all, which is the most prevalent rewarding scheme among consensus mechanisms (*e.g.*, PoW and PoS). But note that this scheme is nonconsequential to PoM, which can be readily extended to allow multiple winners. Realistically, solvers may quit the system because they have outside options. We denote $qc$ as the quitting checkpoint for all solvers: a solver quits, if in $qc$ rounds he or she is not selected even once.

By design, the PoM mechanism would favor solvers with merit, i.e., solvers whose solutions lead to higher sum of users' utility. The system uses a well-defined and commonly accepted metric to measure each user's utility given a solver $k$'s solution, $k \in M_t$. Then, the sum of all users' utility for the solver's solution is calculated, which we use $TU_{k,t}$ to denote. To eliminate low quality solutions, we define a benchmark solution, against which users' solutions are compared (note it is up to the designer to choose the benchmark solution. For example, a simple benchmark could be the average of all feasible solutions proposed or any heuristic that the designer considers as good solution for the problem). The sum of users' utility with the benchmark solution ($TU_{benchmark,t}$) at $t$ is obtained. Then, we calculate Total Utility Gain (TUG) of all solvers and use it as the quality measure of their solutions. To be specific, TUG of a solver $k$ is calculated as the normalized total utility gain of each solver's solution, which is the improvement in the sum of users' utility of the solver's solution relative to that of the benchmark solution. In the case where a solver $k$'s solution is worse than the benchmark solution, i.e., $TU_{k,t} < TU_{benchmark,t}$, his TUG is floored at 0, effectively eliminating low quality solutions. Formally, we define the TUG of solver $k$'s solution at $t$ as follows:

$$TUG_{k,t} = \max\{0, \frac{TU_{k,t} - TU_{benchmark,t}}{TU_{benchmark,t}}\} \quad (1)$$

Note that TUG is the quality measure of a solver's solution based on information available and the quality measuring metric chosen by the system. However, users can reject the high-quality solutions based on constraints which are known only to themselves. Ideally, the evaluation and selection of solutions should take into consideration how users respond to the provided solution. One challenge is that the users' responses to the selected solution can be captured only after the broadcast of the solutions to users.

To address this challenge, we incorporate a post-winning adjustment process in PoM mechanism that uses the acceptance rate to evaluate solvers. We define $\alpha_t = \frac{\text{number of users who accept the solution at period } t}{\text{number of users who receive a solution at period } t}$ as the users' acceptance rate for the winning solution at period $t$. We use $G_{k,t}$ to denote the Winner Selection Score (WSS) of solver $k$ at $t$, which accounts for the $TUG_{k,t}$ (quality) of the solver's solution in this round as well as the users' acceptance rate if she was selected in the previous round. We define post-winning adjustment score of solver $k$ at $t$ as $S_{k,t} = \alpha_{t-1}(1-\lambda)$ if she is the winner in the previous round $t-1$, or $S_{k,t}$ =1, otherwise, where $\lambda \in [0,1]$ is the Decentralized Control Parameter (DCP). We formally define the WSS of solver $k$ at $t$ as follows:

$$G_{k,t} = \sum_{\tau=1}^{t} TUG_{k,\tau} \cdot S_{k,\tau} \quad (2)$$

We illustrate the design of the PoM mechanism in Figure 1. Specifically, at the first matching round ($t = 1$), we initialize the system by setting the post-winning adjustment scores to $S_{k,1} = 1$, for all $k \in M_1$. i.e., each solver's WSS is her TUG score ($G_{k,1} = TUG_{k,1}$) at round 1. In the case where solver $w$ has the highest WSS ($G_{w,1}$ is the highest), she is selected and her solution $X_1^w$ is broadcasted to all the users involved in the solution. Then, the acceptance rate $\alpha_1$ is recorded after the users respond to the solution. Given the value of $\alpha_1$, we adjust the solver $w$'s TUG score in the second solving round by a factor of $S_{w,2} = \alpha_1(1-\lambda)$. It is evident that it is unlikely all users accept the solution ($\alpha_1 < 1$), and thus her WSS will be negatively affected in the second round. On the other hand, the WSS of the other solvers, who did not win in the first round, increases by the amount of their TUG in the second round.

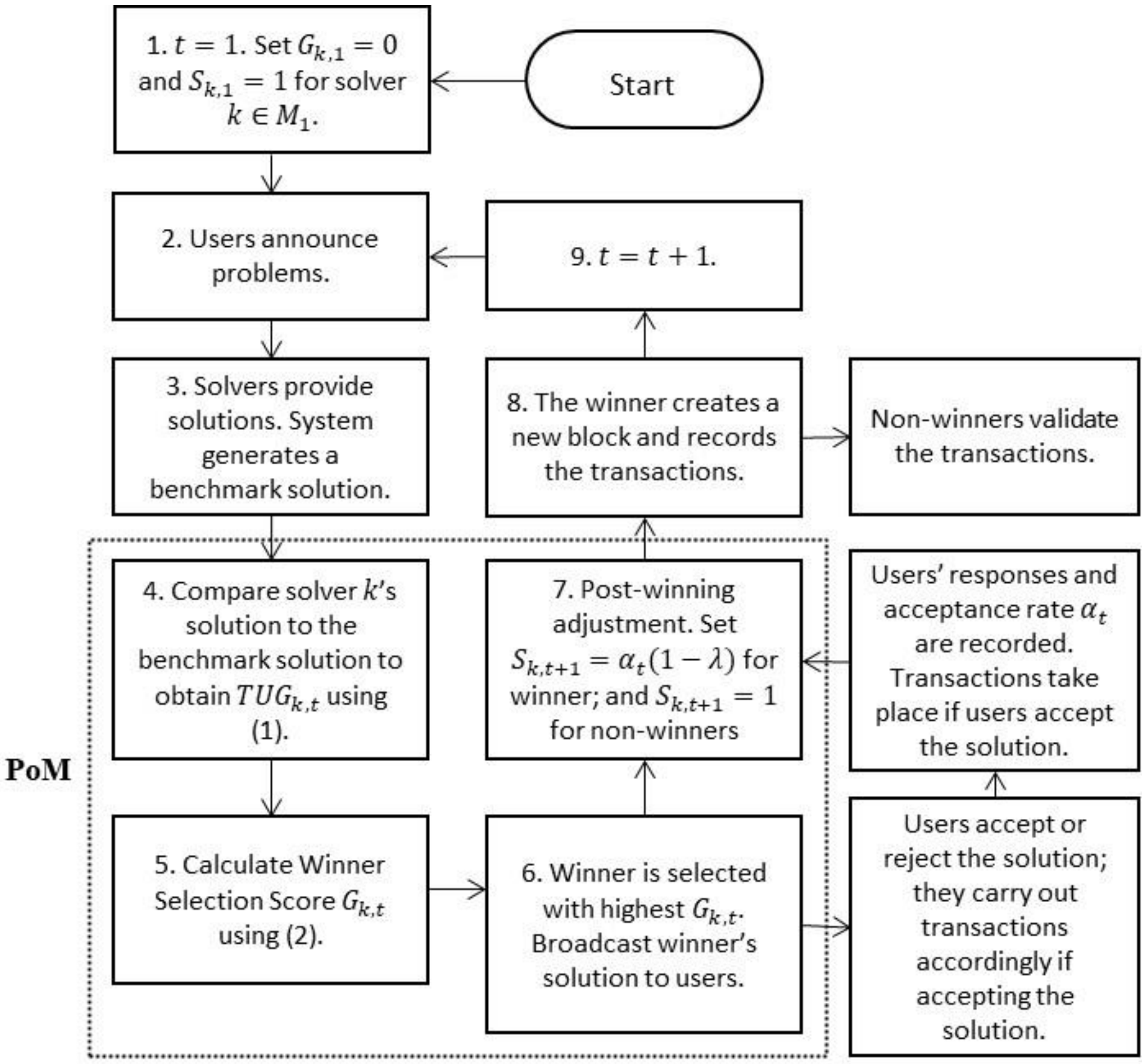

**Figure 1: Flowchart of PoM mechanism**

Note that the acceptance rates are reflections of users' unobservable private constraints. A greater randomness brought by their private constraints will lead to higher randomness in acceptance rates across different periods, which in turn leads to more equitable winning chances among solvers. However, if such randomness is small (*e.g.*, approaching zero), then using the acceptance rates alone cannot effectively prevent a solver from winning consecutively, running into the systematic risk of centralization.

Therefore, the introduction of DCP ($\lambda$) is necessary to prevent the system from degenerating into a centralized one. Without it, it is possible that a small number of solvers - or in the worst case one solver with significant resources - will win all the time, which causes other solvers to quit the system. This parameter $\lambda$ amplifies the impact of the user's acceptance rate on the post-winning adjustment for the selected solver. As $\lambda$ increases, the selected solver's chances of getting selected again in the future rounds decrease, leaving room for the other solvers to win. Note that this selected solver, whose WSS is negatively affected by the post-winning adjustment, can still win in the future rounds. She can continuously provide solutions with high quality (TUG) that boosts her WSS when she does not win, increasing her chances to win in the future rounds . In our conceptualization, the value of DCP ($\lambda$) is controlled by the system designer. This idea of having a designer-controlled parameter to control the system performance is commonly employed in consensus mechanisms (e.g., in Bitcoin's PoW, the designer aims to generate a new block every 10 minutes on average by adjusting the difficulty level of the hashing problem. In Ethereum's PoS, this time window is 14 seconds, where it serves the purpose of design control parameter).

After users respond to (either accept or reject) the winning solution, they carry out transactions according to the solution (only when they accept the solution). The winning solver will create a new block in the blockchain, adding the transactions into the new block and receive a reward for this task. As such, the winner is involved in both the transaction generating and recording processes. For the validation process, we adopt the approach proposed in the PoUW protocol (Turesson *et al*. 2021 and Todorović *et al*. 2022), that is the non-winners in a corresponding round are responsible for validating the transactions in the new block.

## 3.2 Performance of PoM

Under the PoM consensus mechanism, a post-winning adjustment limits the chances of a previous winner succeeding in subsequent rounds, dependent on the acceptance rate of users. This diminishes the likelihood of repeatedly selecting the same high-quality solution provider, promoting fairness within the system—an aspect that signifies PoM's emphasis on *equity*. Meanwhile, the post-winning adjustment needs to ensure that the TUG (quality) of the selected solution does not drastically deviate from the highest-quality one if it isn't selected, representing the *efficiency* aspect of PoM. Next, we formally define these two objective measures.

We consider a measure of inefficiency (inverse of efficiency) of the system at steady state. The steady state reflects solvers' quitting behavior since a solver can exit the system after an extended period of not winning. We denote $s$ as the point that the system achieves a steady state, with the number of solvers remains constant thereafter, (*i.e.*, $m_s \approx m_t, \forall t > s$). After the steady state is reached, we denote the index of the solver who has the highest TUG as $b_t$, and the index of the solver who is selected (has the highest winning selection score) as $w_t$ in round $t$. We define inefficiency as the average difference between the TUG (the efficiency measure does not consider the acceptance rates, since they are assumed to be exogenous and cannot be influenced by DCP in this study) of the winning solution and the highest TUG at the steady state as

$$ine_T = \frac{\sum_{t=s}^{T} TUG_{b_t,t} - TUG_{w_t,t}}{T-s} \tag{3}$$

where $T - s$ is the number of rounds in the steady state.

When a solver continues winning for many rounds, many other solvers may quit because they reach the quitting checkpoint. This leads to a risky situation where the winning solver becomes the only solver available for the remaining rounds. In such a case, she will be selected to provide solutions in all future rounds, which is equivalent to a centralized system. Thus, the PoM mechanism needs to ensure equity to prevent the system from degenerating into a centralized system. We measure equity using the average number of solvers at the steady state, $equ_T$. A higher number of steady-state solvers indicates a more equitable mechanism.

First, at round $t$, we denote $\Gamma_{k,t}$ as the set of periods when solver $k$ was selected as the winner, whereas $\Gamma_{k,t} = \{0\}$ for the solver if she has never won. The set of surviving solvers at period $t$ is

$$M_t = \{k | t - max\, \Gamma_{k,t} \leq qc, \forall k \in M_{t-1}\} \tag{4}$$

where $max\, \Gamma_{k,t}$ is the round of solver $k$'s most recent win at period $t$ (or $max\, \Gamma_{k,t} = 0$ for a solver who has never won). We define the equity of the system as the number of solvers staying in the system at steady state, as

$$equ_T = \sum_{k \in M_s} 1 \tag{5}$$

### 3.3 Application of PoM in a Ride Service On-demand Context

We demonstrate the applicability and universality of the general-purposed PoM mechanism by applying it to a decentralized ride service on-demand setting. Ridesharing platforms (*e.g.*, Uber and Lyft) provide the infrastructure to facilitate "Peer-to-Peer" sharing, allowing drivers to capitalize their spare transportation capacity by offering ride service to passengers. These ridesharing platforms match riders with drivers before riders enjoy ride services (Uber 2024). Recently, several startups have emerged to establish a decentralized ridesharing business model, such as Arcade City, Drife and Aworker. Many essential tasks on these platforms are enabled by blockchain technology. For example, tasks like payment processing and transaction record-keeping are handled by miners on Ethereum at Arcade City. However, these blockchain-enabled ridesharing platforms simply delegate the most critical matching task to individual users themselves: riders list their needs for ride services in an auction and interested drivers bid. Abdulkadiroğlu (2017) questions the efficiency of this self-driven matching mechanism compared to centralized services like Uber's. The question then arises: Can a ridesharing platform use PoM to provide Uber-like matching services, but in a decentralized manner?

Suppose there are riders (users) at a given time on the ridesharing platform. Each rider is interested in getting a ride to their destination location from their current location. Available drivers on the platform at that point have to be matched to a specific rider and if the rider accepts the proposed driver a transaction occurs (for simplicity and without loss of generality, we assume only riders can turn down drivers, but not vice versa). There are $r_t$ riders and $d_t$ drivers ($r_t$ need not to be equal to $d_t$) in a matching round $t$, for $t \in \{1,2,\dots,T\}$. We use $R_t$ to denote the set of riders and $D_t$ the set of drivers at $t$. The platform uses the PoM consensus mechanism to incentivize a group of matchers (solvers) to provide solutions to match riders with drivers and selects one winning solution in every matching round.

A rider's utility depends on the driver who he is matched to (we only consider the rider's utility. However, our model can be easily adjusted to consider the driver's utility). We define rider $i$'s utility of getting a ride from driver $j$ as $u_{ij,t}$ at $t$. The values of $u_{ij,t}$ depend on information available to the solvers and are based on the metric chosen by the platform. For example, the distance between rider *i* and driver *j* can be used to estimate the wait-time to be picked-up (shorter wait-time means higher $u_{ij,t}$ for rider *i* at *t*). The riders may have constraints such as the time by which they need to be at their destination as well as outside options like alternative transportation options, which are known only to them. Naturally, the rider gets zero utility if he is not matched to a driver.

Matchers provide matching solutions, *i.e.*, matching a rider in $R_t$ to a driver in $D_t$ in a one-to-one fashion, so no rider is matched to more than one driver, and no driver is matched to more than one rider. Matchers can provide different matching solutions, which arises from heterogeneity in their innate abilities to solve complex problems, such as their use of different heuristics for the matching problem (Fisher *et al*. 1986; Savelsbergh 1997). If a matcher $k$ matches rider $i$ to driver $j$, then $x_{ij,t}^{k} = 1$; or $x_{ij,t}^{k} = 0$, otherwise, where $x_{ij,t}^{k} \in X_t^k$ for all $i \in R_t$ and $j \in D_t$. The sum of riders' utility given matcher *k*'s solution set at *t* is $\sum_{j \in D_t} \sum_{i \in R_t} \left( u_{ij,t} \cdot x_{ij,t}^{k} | U_t, X_t^k \right)$. Although the merit of the solution can be judged on different dimensions, *e.g.*, wait time for riders and/or cost to drivers, for illustration purposes, we use the waiting time for riders.

We use $X_t^0$ to denote the benchmark solution where riders and drivers are randomly paired at *t*. The sum of riders' utility with the benchmark solution at *t* is $\sum_{j \in D_t} \sum_{i \in R_t} \left( u_{ij,t} \cdot x_{ij,t}^{0} | U_t, X_t^0 \right)$. Under the ridesharing context, we calculate *Total Utility Gain* of matcher *k*'s solution at *t* as:

$$TUG_{k,t} = \max\{0, \frac{\sum_{j \in D_t} \sum_{i \in R_t} \left( u_{ij,t} \cdot (x_{ij,t}^{k} - x_{ij,t}^{0}) | U_t, X_t^k, X_t^0 \right)}{\sum_{j \in D_t} \sum_{i \in R_t} \left( u_{ij,t} \cdot x_{ij,t}^{0} | U_t, X_t^0 \right)}\} \quad (6)$$

After receiving a proposed match, a rider may choose to not accept it due to her private constraints (for simplicity, we assume that only riders reject or accept the matches. This mimics the current Uber type of business model where the rider decides on whether to choose the driver proposed to be matched to him. But our model is insensitive to which side (e.g., driver) makes the acceptance decision). For example, she could reject the proposed match if the wait-time for pick-up exceeds that of her alternative transportation option, such as taking a bus or walking to her destination. The platform does not have the knowledge about the riders' private constraints, but only observes ex-post individual riders' responses to the winning matching solution. The platform records the acceptance rate at $t$, i.e., $\alpha_t = \frac{\text{number of riders who accept the match at period } t}{\text{number of riders who receive a match at period } t}$, which will be used to calculate $S_{k,t}$ (post-winning adjustment score) of all solvers, affecting their $G_{k,t+1}$ (winner selection score) at round $t$+1.

In the case where a rider accepts the matching solution, the information of the rider and the matched driver will be exchanged between the two. Subsequently, the drivers deliver the ride service to the riders according to the solution, and transactions take place. The winning matcher registers the transactions into a new block in the platform's blockchain and receives a reward from the platform. The matched riders and the drivers will then be removed from sets $R_t$ and $D_t$, respectively. In the case where a rider rejects the proposed match, he can opt to wait for the next matching round or leave the platform altogether. At the conclusion of each matching round, the platform accommodates new drivers and riders.

# 4 Agent Based Simulation

In this section, we use ABS to explore the properties and performance of the PoM mechanism (using Python's Mesa package). The *first objective* of the simulation is to study the impact of the system design parameter, DCP ($\lambda$), on the efficiency and equity aspects of PoM in the steady state, under various environmental conditions such as the level of skill heterogeneity and quitting behavior of matchers. In line with the ABS framework laid out in the literature (Chan and Gao 2013; Nan 2011; Malgonde *et al*., 2020; Zhang *et al.*, 2020a; Prawesh and Padmanabhan 2021), we set up the simulation in Section 4.1 and define the agents' attributes accordingly: (1) the environment attributes, and then (2) the interactions between the agents and the environment.

We simulate agents as a group of matchers who provide matching solutions, e.g., matching riders and drivers on a ridesharing platform, with varying qualities in each iteration. We experiment with different distributions of matcher capability to study the impact of DCP on the efficiency and equity of PoM under different degrees of skill heterogeneity of matchers. For simplicity, we use uniform distribution to capture the heterogeneity of matchers. We also specify the number of matchers who provide solutions in the system at the beginning. The environment is configured according to the PoM consensus mechanism described in Section 3, including $\lambda$ (DCP), $T$ (total number of periods), $\alpha$ (acceptance rate), and $qc$ (quitting checkpoint).

First, $T$ is set up to be sufficiently large for the system to achieve a steady state. We use different values of DCP to study its impact. The expectation is that high DCP will result in PoM sacrificing efficiency to improve equity, while low DCP will compromise equity to enhance efficiency. The agents interact with the environment in three ways: (1) matchers provide matching solutions; (2) users accept or reject these proposed solutions; and (3) matchers decide whether to quit or stay in the system after the winner is selected. All the interactions follow the winner selection process detailed in Section 3.2. We summarize and discuss the results in Section 4.2.

Our *second objective* of the simulation is to demonstrate how a system designer, given their preferences, can identify desirable tradeoffs between equity and efficiency by appropriately choosing DCP. In Section 4.3, we first describe the characteristic of the problem and then locate the appropriate DCP for the system designer through simulation.

## 4.1 Simulation Setup

We simulate 1,500 matching rounds in each epoch, and we simulate 500 epochs to smooth out the distributions considered in our context. Our preliminary analysis indicates that on average it takes 1,200 rounds for the system to enter a steady state, therefore 1,500 matching rounds is sufficient for having $T-s$ periods after steady state to analyze inefficiency. We assume there are 100 heterogeneous matchers to start with (we use 100 to be close to Bitcoin's requirement of having at least 101 miners), each of whom has a different capability in producing matching solutions in each epoch. We denote the capability of matcher $k$ as $q_k, k \in M_0$, and a priori quality of the matching solution that matcher $k$ proposes as $TUG_{k,t}$ at round $t$. At the beginning ($t=0$) of each epoch, each matcher's capability $q_k, k \in M_0$, is drawn from a common uniform distribution, which is $U(L,1)$ where $L \in (0,1)$. For each subsequent round, the priori quality of the matching solution of matcher $k$ is equal to her capability (*i.e.*, $TUG_{k,t} = q_k$). At round 0 of the simulation run, matcher $k$'s initial WSS, $G_{k,0}, k \in M_0$, is set to 0. We update the WSS of each matcher by adding the adjusted scores $TUG_{k,t} \cdot S_{k,t}$ into $G_{k,t}$ (Equation 2) and decide the winner as the one with the highest $G_{k,t}$ at round $t$.

We vary the environment to simulate different conditions. We simulate changes in the matchers' capability distribution by setting different lower bound values (*i.e.*, $L$ of the uniform distribution) at $0.9, 0.7, 0.5$ and 0.3. For simplicity of the simulation, we only vary the capability distribution. The main purpose is to show that PoM mechanism does not result in centralization under different environments. Varying other parameters such as acceptance rate does not alter this qualitative result. We choose those values so that the results are representative (i.e., do not fall into extreme values) but are distinguishable (i.e., having large enough interval to ensure heterogenous capability). We also simulate different values of $\lambda$ over a 0.1 interval within 0 to 1 (we also experimented with smaller intervals, but they do not provide additional insights from our current results). Given $\alpha_t$, the next round adjustment score for the winner of round $t$ is $S_{w_t,t} = \alpha_t(1 - \lambda)$ and for all other matchers, $S_{w_t^-,t} = 1$ (we choose $\alpha_t = 0.8$ as exogenously given, since users have private information that is unobservable. We find that different values of $\alpha_t$ do not provide additional insights to our analysis). We calculate the average speed of convergence, the average steady state *inefficiency* and *equity* in terms of *average number of steady-state matchers*, number of wins for each matcher, and average Gini coefficient over 500 epochs. We count the number of times that each matcher's solution is selected in the steady state and calculate the Gini coefficient based on Dorfman (1979). Note that the Gini coefficient and the number of wins of each matcher capture different aspects of equity, but the results are largely consistent. Then we construct the $ROC$ curve to reveal the tradeoff between efficiency and equity based on *inefficiency* and *average number of steady state matchers* at different values of $\lambda$. Finally, we investigate the appropriate choice of DCP for a decentralized platform designer given her targeted outcomes.

## 4.2 Simulation Results

### 4.2.1 Effect of DCP and capability distribution on efficiency

Not surprisingly, our simulation results suggest that higher DCP increases inefficiency, or in other words reduces efficiency. We take the average of inefficiency in the steady state across all 500 epochs, given the same matcher distribution and DCP. Figure 2(a) and Figure 2(b) show that higher DCP increases the mean and variance of inefficiency in the steady state.

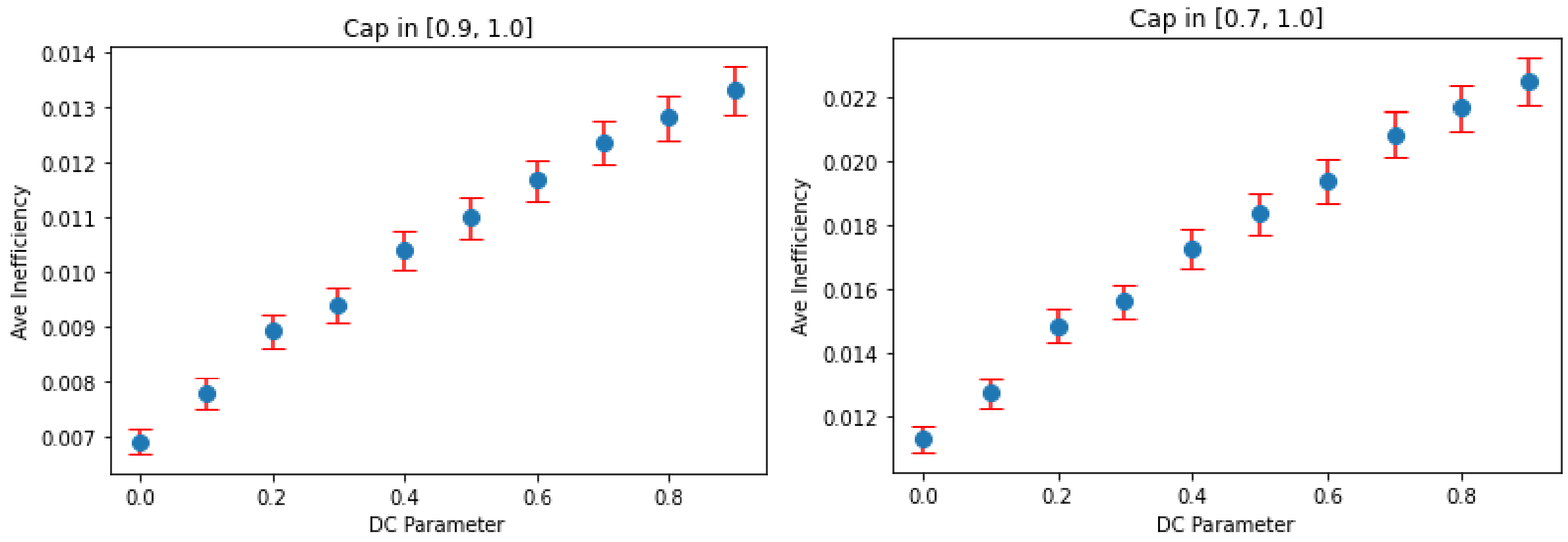

(a) matcher distribution [0.9, 1.0]  (b) matcher distribution [0.7, 1.0]

**Figure 2: Mean and standard deviation of inefficiency**

In both figures, the *x*-axis represents DCP where $\lambda$ ranges from 0 to 0.9, while the *y*-axis indexes the absolute value of inefficiency. We do not consider a $\lambda$ between 0.9 to 1 because value that is too large (close to 1) will effectively make winning selection a random process instead of a merit-based scheme. Figures 2(a) and 2(b) differ in terms of the matcher-capability distribution. The former is drawn from a uniform distribution of [0.9,1] that represents the case of having a group of similarly capable matchers, meanwhile the latter is drawn from [0.7, 1] that represents a more heterogenous group of matchers from the start. Nonetheless, the trends across the two figures turn out to be the same - inefficiency increases as $\lambda$ goes higher. Moreover, when there is a higher level of heterogeneity present in matchers' capabilities (Figure 2(a)), both the average and standard deviation of inefficiency increase. This result shows that inefficiency increases when the difference among matchers' capabilities is high, even though the winning probabilities of low capability matchers becomes lower.

In addition, we find that a higher DCP reduces the number of periods it takes to converge to steady state, as shown in Figure 3. The reason is that with a higher DCP value, the winners' winning probabilities in the future are subject to a bigger downward adjustment. However, matchers who constantly provide high-quality solutions can regain their advantage through accumulating WSS over time more easily. In contrast, it becomes harder for matchers who have relatively low capability to keep winning with pure "luck". Thus, it takes a fewer number of periods before these low-capability matchers quit as compared to a system with a small DCP value.

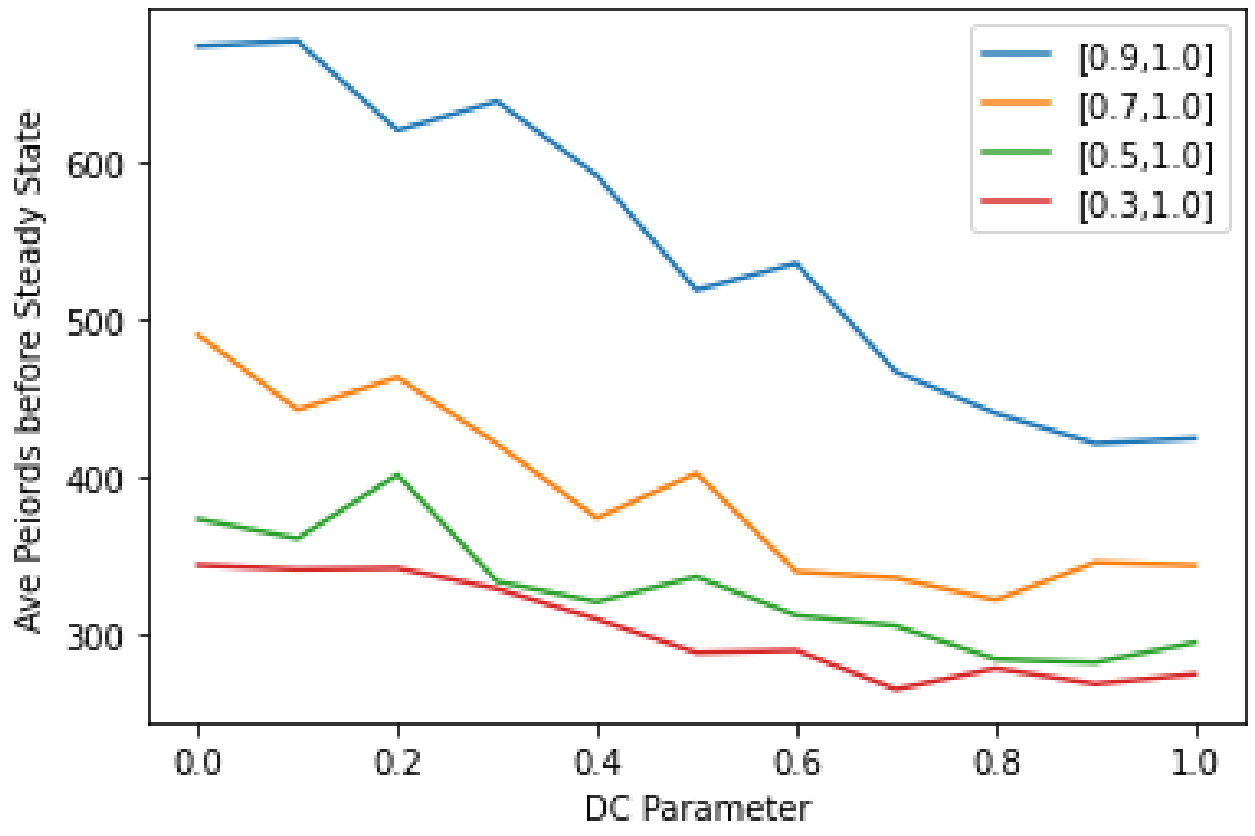

**Figure 3: Number of periods before steady state**

When there exists a higher level of heterogeneity in matchers' capabilities (*i.e.*, matcher distribution is on the support [0.5, 1.0] and [0.3, 1.0]), it takes a smaller number of periods before the PoM system enters the steady state. Intuitively, in the high heterogeneity scenario, matchers with low capabilities have less chance to compete with others, since the sheer increment speed of their WSS is far lower than those with higher quality of solutions on average. Thus, matchers with relatively low capabilities quit sooner, as compared to a system with a low level of capability heterogeneity.

### 4.2.2 Effect of DCP and capability distribution on equity

Next, we look at how DCP and matcher distribution affect equity. Our results suggest that when the DCP turns higher, more matchers will survive on the platform in the steady state, thus higher equity as shown in Figure 4. Moreover, the number of surviving matchers goes down as the heterogeneity in matcher capability becomes higher, which suggests a lower equity level.

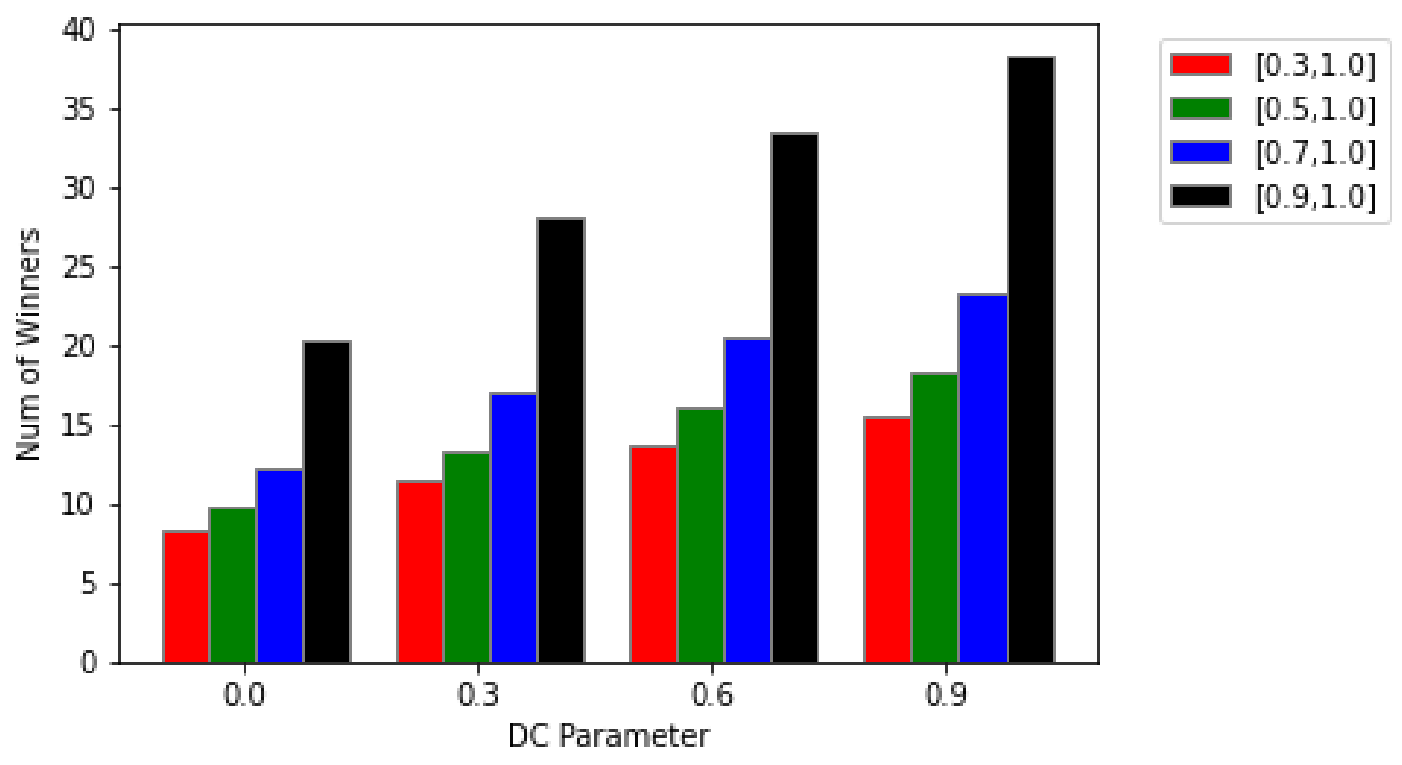

**Figure 4: Number of matchers in the steady state**

In Figure 4, we plot the average number of wins ($y$-axis) for matchers on the $x$-axis that indexes their relative capability percentiles. For example, when the support of matcher distribution is [0.9, 1.0], a matcher $k$ with capability $q_k = 0.98$ has capability percentile 0.8 on the $x$-axis, since there are 80% matchers having capabilities lower than her. Similarly, when the support of a matcher distribution is [0.7,1.0], a matcher $k$ with capability $q_k = 0.94$ also has capability percentile 0.8 on the $x$-axis because 80% matchers fall into the capability range [0.7,0.94].

As shown in Figure 5(a), when there is less heterogeneity (*e.g.*, lower bound of capability distribution being 0.9 instead of 0.7), matchers with high capability percentiles experience a smaller number of wins. For instance, in Figure 5(a), when the support of matcher distribution is [0.9,1.0], a matcher with capability percentile 1.0, who provides solution with the highest quality, reaps 70 wins on average; but when the support becomes [0.7,1.0], a matcher with the same percentile can gather 140 wins on average.

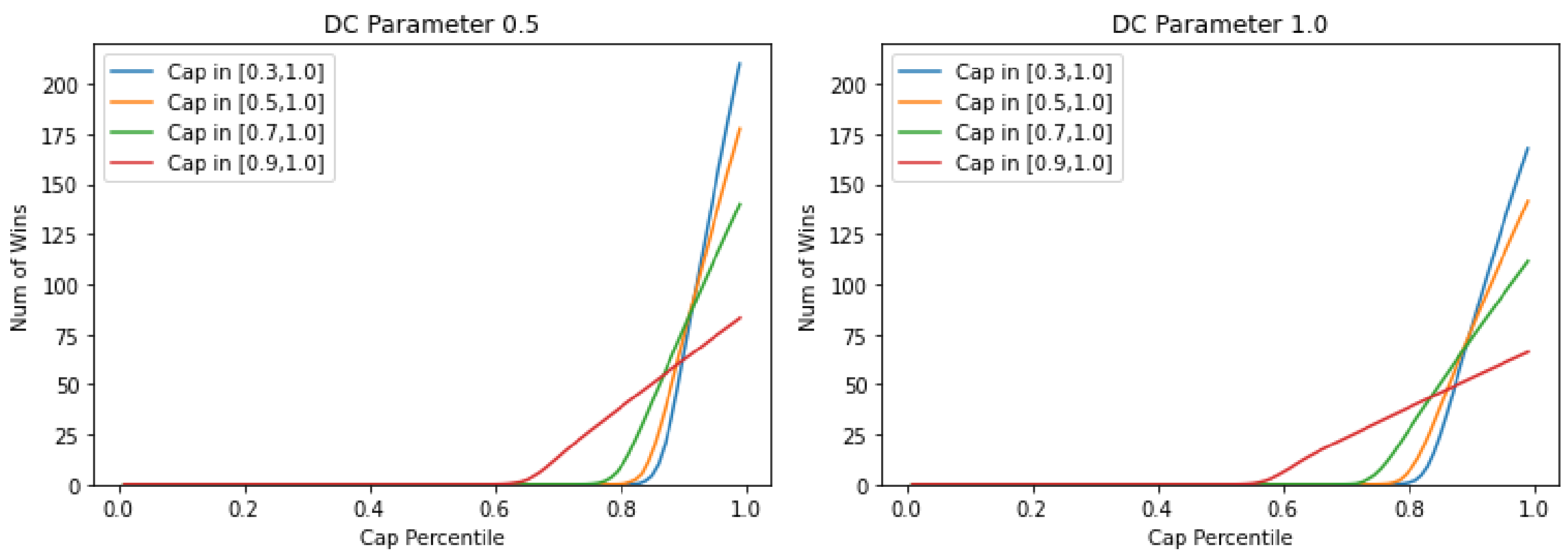

(a) DCP = 0.5 (b) DCP = 1.0

**Figure 5: Number of wins as a function of DCP**

The reason is that, under a capability distribution where matchers are more homogeneous (*i.e.*, the case of [0.9,1.0]), lower-capability matchers are not too different from those with higher capability. As a result, even those matchers with relatively lower capabilities still have a significant chance to win. So, their number of wins does not fall too much behind those of higher-capability matchers. Conversely, if matcher capabilities are more heterogeneous (*i.e.*, in the case of having a support of [0.7,1.0]), the lower-capability matchers are far less capable of producing high quality solutions. Consequently, lower-capability matchers earn less chance to win and most of the number of wins will instead go to the hands of higher-capability matchers.

When DCP increases (Figure 5(b)), the wins move away from the hands of high-capability matchers to the hands of low-capability matchers. Because high-capability matchers tend to win more frequently in the case of a low DCP, the amount of downward adjustment they will receive becomes higher, which then lowers their winning probabilities.

Lastly, we treat the number of wins of each matcher as their wealth and measure the equity of wealth distribution using the Gini coefficient. The Gini coefficient calculation follows the method as defined in Dorfman (1979) by replacing income with the number of wins of a matcher.

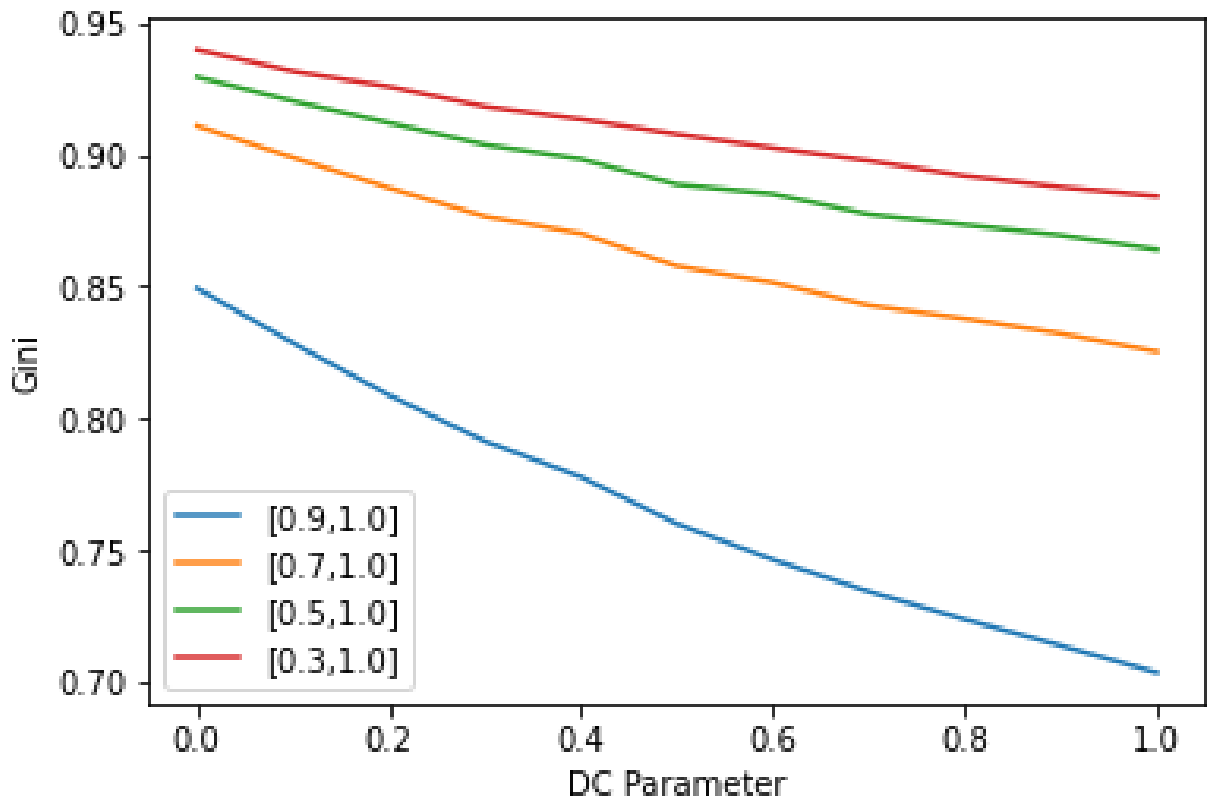

**Figure 6: Gini coefficient**

Figure 6 suggests that the Gini coefficient is smaller (thus, higher equity) when the DCP value goes higher; further the Gini coefficient becomes smaller for more homogeneous matchers.

#### 4.2.3 Tradeoff between efficiency and equity under different DCP values

Evidently there exists a tradeoff between equity and efficiency when the DCP value changes. We graphically highlight this tradeoff using a ROC type of curve as shown in Figure 7. We plot the average steady state inefficiency on the x-axis and the average *number of steady state matchers* (equity) on the y-axis in Figure 7. The scales of both the x-axis and y-axis are normalized to be between 0 and 1 (*i.e.*, 100% on the x-axis (y-axis)) represents the maximum level of inefficiency (equity), and 0% represents the minimum level of inefficiency (equity) that can be obtained by applying the DCP to be 1 and 0, respectively.

From the curve, we can derive the inefficiency and equity at different values of $\lambda$, given a certain set of exogenous parameters $\Theta$. We use $ine_T(\lambda, \Theta)$ to represent the steady state inefficiency given $\lambda$ and $\Theta$, and use $equ_T(\lambda, \Theta)$ to represent the steady state equity given $\lambda$ and $\Theta$, where $T$ is the total number of periods in a simulation. It can be seen that the curve is close to a straight line, meaning that it is not possible to improve equity without compromising inefficiency.

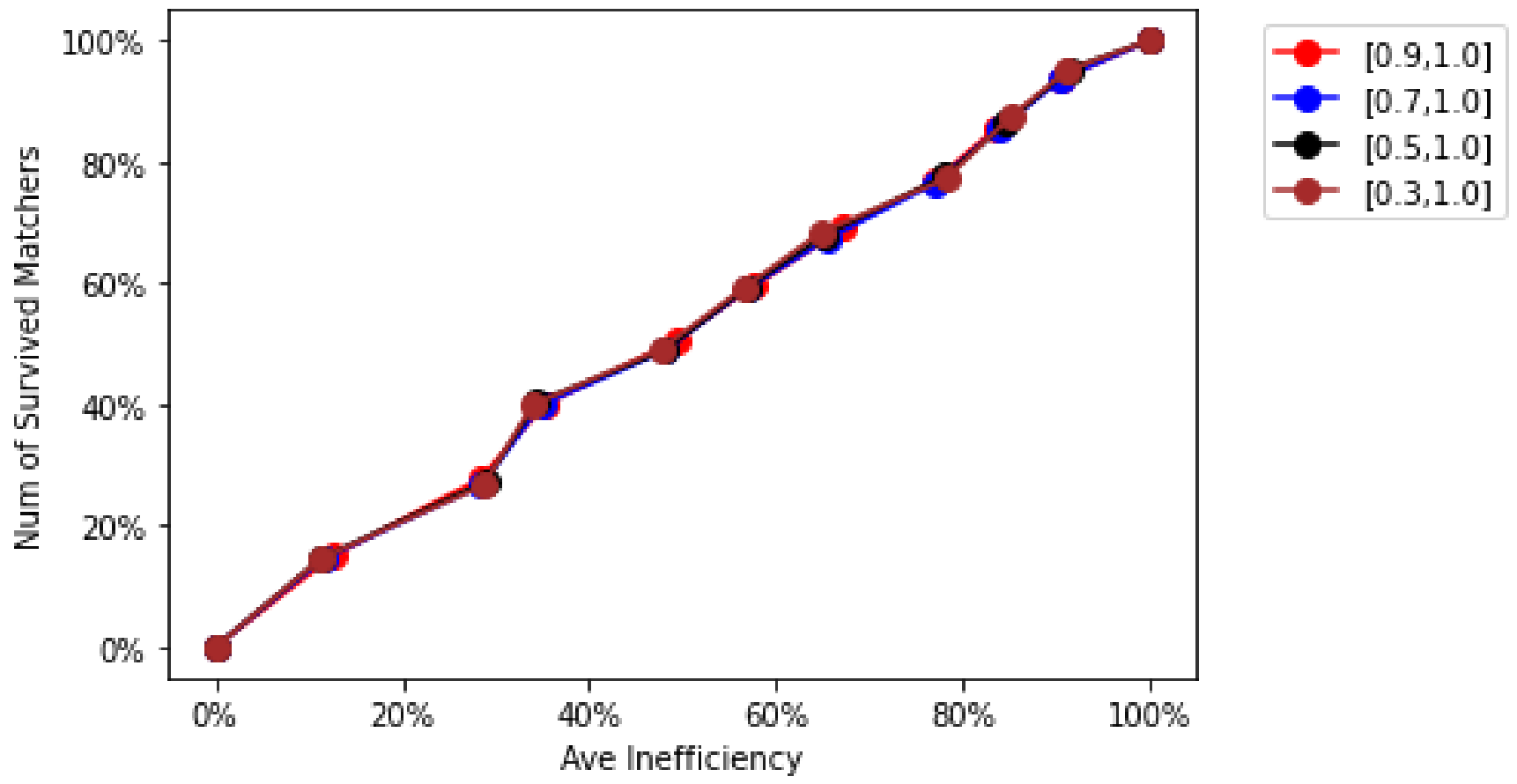

**Figure 7: ROC-type of curve (normalized)**

#### 4.2.4 Discussion on the results

In summary, the simulation yields several key findings on DCP's impact. First, with respect to efficiency, increased DCP heightens both the average inefficiency and the variance of the winning solution. A rise in DCP from 0 to 1 almost doubles the average inefficiency, as high-quality solution providers win less frequently due to the amplified post-winning adjustment, leading to increased average inefficiency and variance.

Second, regarding equity, elevated DCP enhances the average equity of the PoM system. A surge in DCP from 0 to 1 nearly doubles the number of matchers remaining in in the steady state, as higher DCP results in a steeper drop in a winner's future winning probability. Consequently, more low-quality solution providers remain in the system, increasing the system's average equity (Figure 5).

Third, these findings suggest an absolute tradeoff between efficiency and equity in a PoM system, making it impossible to achieve high levels of both concurrently. Deciding on an appropriate DCP level, a crucial question for a PoM system designer we discuss next.

## 4.3 Problem Characteristics and DCP Choice

DCP is a key design parameter of the PoM consensus mechanism. It is evident from the simulation that the value of DCP directly influences the equity and efficiency tradeoff. An increased DCP lowers the likelihood of the winner being selected again in future periods due to a decrease in her WSS. This improves equity by allowing other matchers to win. Naturally, from a system design point of view, the choice of DCP should be made in a such a way that maintains a desirable level of equity without significantly compromising efficiency.

Analytically, solving the DCP optimization problem is challenging because of: (1) the exogenous nature of riders' acceptance rate; (2) the complication that the set of matchers remaining on the platform is different at different matching rounds; and (3) the complexity of matchers' quitting behavior. Therefore, we continue to rely on ABS to discern the appropriate DCP value. In practice, system designers may assign different weights to competing objectives (Yu *et al*. 2019). Assuming the system designer's preference type is $\beta, \beta \in (0,1)$, where a larger $\beta$ indicates a stronger preference for equity. We employ a Cobb-Douglas type of function to account for the diminishing effect of higher equity or higher efficiency on a system designer's objective. Consequently, the DCP choice problem is formulated as follows:

$$\max_{\lambda} \left(\frac{equ_T(\lambda,\Theta)-equ_T(0,\Theta)}{equ_T(1,\Theta)-equ_T(0,\Theta)}\right)^{\beta} \cdot \left(\frac{ine_T(1,\Theta)-ine_T(\lambda,\Theta)}{ine_T(1,\Theta)-ine_T(0,\Theta)}\right)^{1-\beta} \tag{7}$$

subjects to,

$$equ_T(\lambda, \Theta) \geq equ_{min} \tag{8}$$

$$ine_{max} \geq ine_T(\lambda, \Theta) \tag{9}$$

$$1 \geq \lambda \geq 0 \tag{10}$$

where $equ_{min}$ is the minimum required number of steady state matchers (equity) and $ine_{max}$ is the maximum allowed level of inefficiency. The parameter $\Theta$ denotes the set of all exogenous parameters (*e.g.*, acceptance rate) besides $\lambda$. In (7), the denominators are the maximum level of equity (inefficiency) minus the minimum level of equity (inefficiency) when the DCP takes a value of 1 and 0, respectively. Equations (8), (9) and (10) are the constraints of this problem. Next, we use an example to illustrate the DCP choice problem.

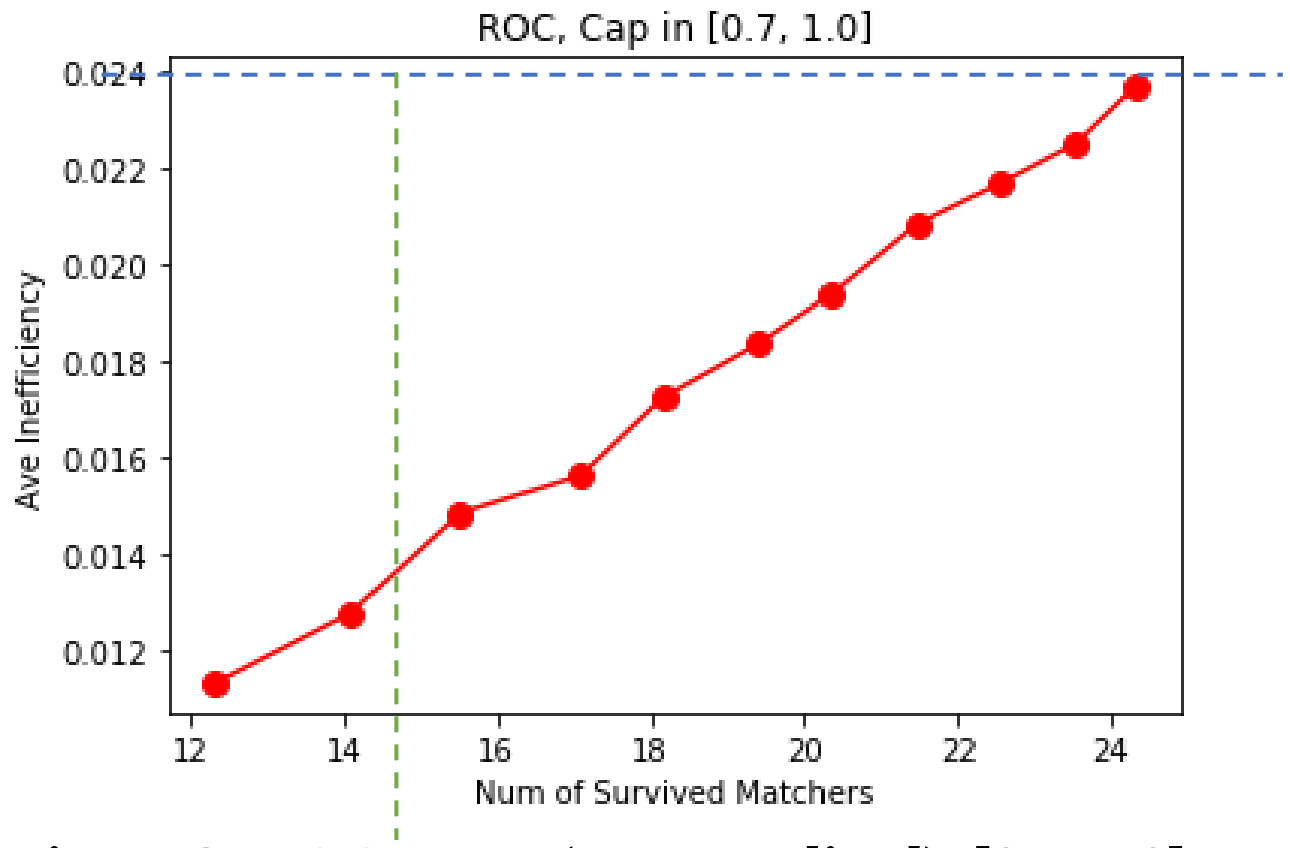

**Figure 8: ROC curve (unnormalized), [0.7, 1.0]**

In the example, we assume the support of the distribution of matchers is [0.7, 1.0]. Figure 7 shows that the objective function is invariant to our choice of distribution of matchers. And the constraints are the following: $equ_{min} = 15$ and $ine_{max} = 0.025$. That is, the system designer wants to make sure that the choice of DCP guarantees at least 15 matchers remaining in the steady state on average, with at most an average inefficiency of 0.025 in the steady state. Figure 8 shows that, after considering the constraints, the feasible choices of DCP are $\lambda \in \{0.2, 0.3, 0.4, 0.5, 0.6, 0.7, 0.8, 0.9, 1.0\}$. For any $\beta$, the choice of $\lambda$ is the one that maximizes objective function (7). The appropriate choice of $\lambda$ and the objective function values for different levels of $\beta$ are shown in Table 1. The value of $\lambda$ increases as the system designer's type $\beta$ increases, where larger $\beta$ indicates that the equity is weighted more.

**Table 1: Choice of DCP, [0.7, 1.0]**

| **Designer's type $\beta$** | **Choice of $\lambda$** | **Objective function values** |
|---|---|---|
| $\beta \in \{0.1, 0.2\}$ | $\lambda$ =0.2 | {0.65, 0.59, 0.55} |
| $\beta \in \{0.3, 0.4, 0.5\}$ | $\lambda$ =0.3 | {0.56, 0.54, 0.51} |
| $\beta = 0.6$ | $\lambda$ =0.5 | 0.52 |
| $\beta = 0.7$ | $\lambda$ =0.6 | 0.55 |
| $\beta = 0.8$ | $\lambda$ =0.8 | 0.61 |
| $\beta = 0.9$ | $\lambda$ =0.9 | 0.74 |

### 4.3.1 Comparing different matcher distributions

We compare the choices of DCP under different matcher distributions. For the same set of constraints ($equ_{min} = 15$ and $ine_{max} = 0.025$), the feasible solution set of $\lambda$ is larger for a more homogeneous distribution of matcher capabilities as shown in Figure 9.

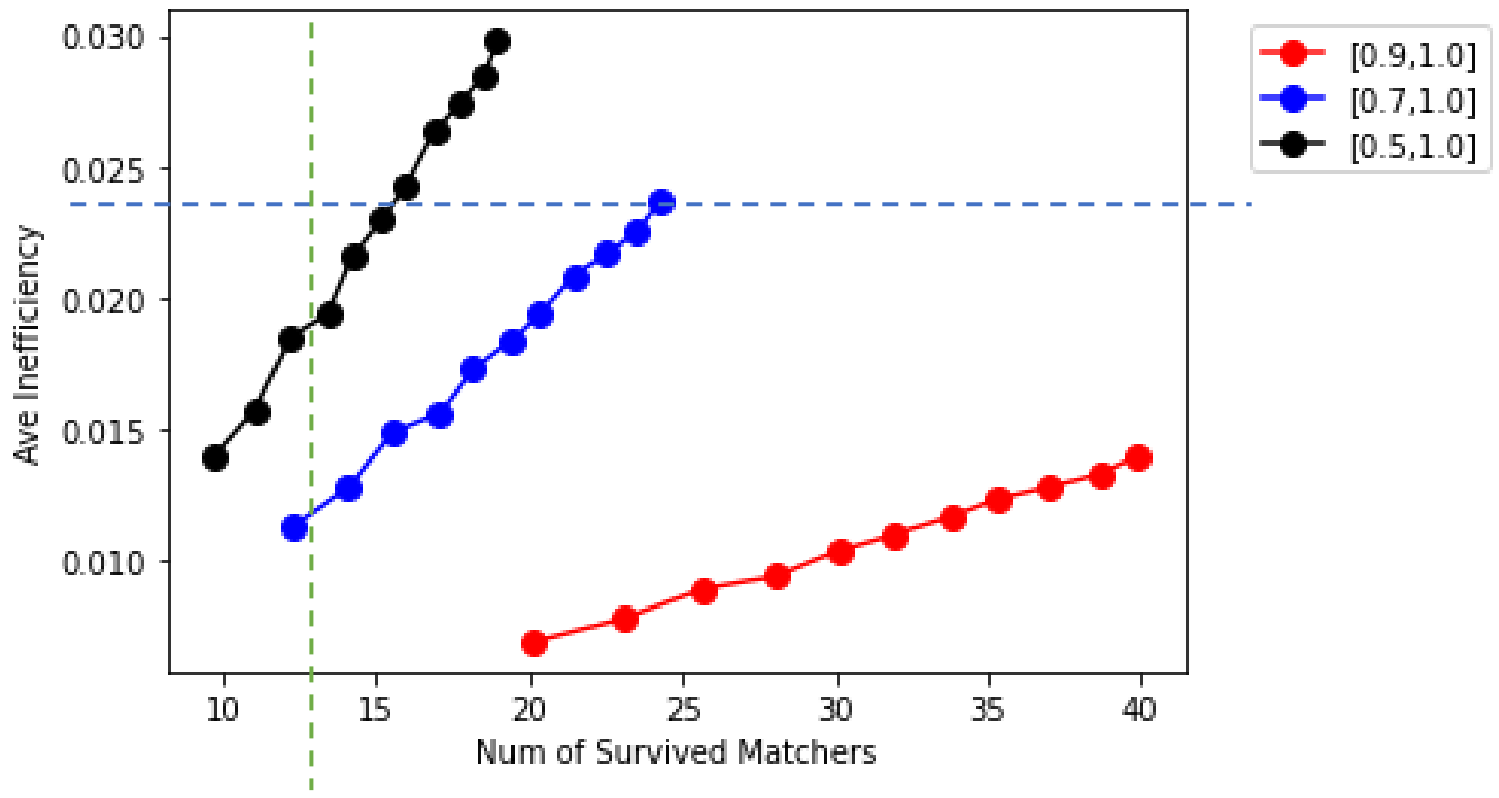

**Figure 9: ROC curves (unnormalized)**

**Table 2: Choice of DCP, [0.9, 1.0]**

| **Designer's type $\beta$** | **Choice of $\lambda$** | **Objective function values** |
|---|---|---|
| $\beta \in \{0.1, 0.2\}$ | $\lambda$ =0.1 | {0.73, 0.61} |
| $\beta \in \{0.3, 0.4, 0.5\}$ | $\lambda$ =0.3 | {0.56, 0.53, 0.51} |
| $\beta = 0.6$ | $\lambda$ =0.5 | 0.52 |
| $\beta = 0.7$ | $\lambda$ =0.6 | 0.55 |
| $\beta = 0.8$ | $\lambda$ =0.8 | 0.61 |
| $\beta = 0.9$ | $\lambda$ =0.9 | 0.74 |

Table 2 shows the DCP choice when the matcher capability distribution is on the support [0.9,1.0], representing homogenous matchers. In this case more feasible DCP choices become possible. Further, the appropriate choices can be different under the same $\beta$ as compared to distribution [0.7,1.0]. For example, the DCP choice of 0.1 is infeasible under the matcher distribution of [0.7,1.0], but it becomes feasible under the distribution of [0.9,1.0].

In Table 3, we also display the appropriate DCP choices when the capability distribution is on the support [0.5,1.0], where the system designer has fewer choices of $\lambda$. Our simulation results suggest that the system designer should choose different DCPs to maximize the objective by considering both the designer's own preferences on equity and efficiency, and the distribution of matchers available in the system.

**Table 3: Choice of DCP, [0.5, 1.0]**

| Designer's type $\beta$ | Choice of $\lambda$ | Objective function values |
|---|---|---|
| $\beta \in \{0.1,0.2,0.3,0.4,0.5,0.6\}$ | $\lambda$ =0.5 | $\{0.44,0.46,0.47,0.49,0.51,0.52\}$ |
| $\beta \in \{0.7,0.8,0.9\}$ | $\lambda$ =0.6 | $\{0.56, 0.59, 0.63\}$ |

## 5. Conclusion and Discussion

This paper introduces a novel *generative blockchain* framework, seeking to integrate the transaction generation process into the existing transaction recording process facilitated by conventional blockchain systems. Towards this end, we propose a general-purpose blockchain consensus mechanism, Proof-of-Merit (PoM). PoM solicits solvers to solve users' complex problems related to transaction generation and selects a winning solution based on merit. The selected solution governs generating and recording of these transactions onto the blockchain. In contrast, existing consensus mechanisms, such as PoW, PoS and PoUW, are used to record existing transactions but not generate new transactions. Even when the computing power is being used for solving a problem, as in the case of PoUW, the problem is being solved for an external party rather than the problem at hand.

Furthermore, PoM effectively balances efficiency and equity. It factors in the historical quality of each solver's contributions when choosing the winner, thus entailing a merit-based system; it assures equity by leveraging Dynamic Capability Penalty (DCP) to prevent dominant solvers from monopolizing the wins, preserving the decentralization principle. The implications of varying DCP on PoM's performance were explored through agent-based simulation, demonstrating the inevitable trade-off between efficiency and equity. Our simulation also established the appropriate DCP level from a system designer's perspective, revealing that the desired balance between equity and efficiency can be contingent on various circumstances.

Our study contributes to blockchain literature. First, our proposed framework of generative blockchain has the potential to transform the blockchain industry, by switching the focus from transaction recording to transaction generation in complex environments. Second, we devise a new consensus mechanism to decentralize the task of transaction generation. In so doing, we allow solvers to solve a meaningful social-benefiting (business) problem as opposed to miners who only solve an energy-wasting puzzle. PoM goes beyond the extant consensus mechanism from record-keeping of the existing transactions to providing a solution that allows for new transactions to be created. This represents a significant leap in the design of consensus mechanisms. Third, our PoM mechanism, as a merit-based system, determines the winner based on the quality of the solution a solver provides. As such, it diverts the precious computing resources towards providing high-quality solutions, thus improving sustainability of blockchain. Fourth, our use of ABS offers several new insights on the performance and properties of PoM. For example, it answers “what-if” questions, such as what proper levels of DCP should one choose to best balance efficiency with equity under a specific business environment.

We also provide managerial implications associated with the proposed mechanism., we first emphasize the necessity and benefit of adopting a mechanism in businesses where the solutions to complex problems govern what transactions are to be generated and recorded. When a business requires solving complex business problems in generating a transaction, combining the consensus process with the transaction generation process is not only feasible, but it may also be necessary from a sustainability point of view. The computational cost of the PoW consensus mechanism can become unsustainable, especially for its excessive consumption of energy. Second, a system control parameter like DCP prevents the decentralized system from falling into the trap of centralization, and it improves the overall equity while not overly compromising on efficiency. When developing such a mechanism, a blockchain system designer should not only consider the security and speed aspects (as in the design of PoW and PoS), but also the equity and efficiency aspects of the system. Our work reveals that there exists an absolute tradeoff between equity and efficiency. Therefore, a system designer cannot achieve efficiency and equity at the same time. One may need to sacrifice efficiency to achieve equity; for example, where the number of problem solvers is small or where the skills of these solvers vary greatly. However, if enough number of problem solvers are available or the skills of the problem solvers are similar, a system designer can afford to trade equity for efficiency.

There are several possible extensions to the current work. First, in the current setup, we endorse a "winner-takes-all" rewarding mechanism where only one winner is selected. To encourage more solvers to participate, allowing more winners might be a conceivable solution. Such a multiple-winner rewarding mechanism has been recently proposed in the blockchain world, referred to as the Ommer block (Investopedia 2024). Second, we assume that the acceptance rate comes from riders in an exogenous manner. Endogenizing this relationship and studying its impact on the performance of PoM would warrant an interesting future study. Third, it would be interesting to study the performance of PoM when existing solvers could quit, but at the same time new solvers can be attracted to join.